\documentclass[useAMS,usenatbib]{mn2e}

\usepackage{natbib}
\usepackage{graphics}
\usepackage{array}
\usepackage{stfloats}
\usepackage{fixltx2e}
\usepackage{amsmath}
\usepackage{fontenc}
\usepackage{amssymb,amsfonts,txfonts}
\usepackage{threeparttable}
\usepackage{longtable}
\usepackage{graphicx}
\usepackage{url}
\usepackage{textcomp}
\usepackage{color}
\usepackage{times}
\fontencoding{T1}

\newif\ifAMStwofonts
\AMStwofontstrue

\newcommand{\be}{\begin{equation}}
\newcommand{\ee}{\end{equation}}
\newcommand{\bea}{\begin{eqnarray}}
\newcommand{\eea}{\end{eqnarray}}
\newcommand{\fnl}{f_{\mathrm{NL}}}
\newcommand{\qnl}{q_{\mathrm{NL}}}
\newcommand{\gal}{{\mathrm{gal}}}
\newcommand{\obs}{{\mathrm{obs}}}
\newcommand{\omegacdm}{\Omega_{\mathrm{c}}}
\newcommand{\omegam}{\Omega_{\mathrm{m}}}

\renewcommand{\vec}[1]{ {\bmath #1} } 

\def\ltsima{$\; \buildrel < \over \sim \;$}
\def\simlt{\lower.5ex\hbox{\ltsima}}
\def\gtsima{$\; \buildrel > \over \sim \;$}
\def\simgt{\lower.5ex\hbox{\gtsima}}

\title[Clustering and abundances of galaxy clusters]{Combining clustering and abundances of galaxy clusters\\to test cosmology and primordial non-Gaussianity}
\author[A. Mana, et al.]{Annalisa~Mana$^{1,2}$, Tommaso~Giannantonio$^{1,2}$,
 Jochen~Weller$^{1,2,3}$, Ben Hoyle$^{1,2}$, \newauthor
 Gert~H\"utsi$^{4,5}$ and Barbara~Sartoris$^{1,2,6,7}$\\
$^1$ Universit\"ats-Sternwarte M\"unchen, Ludwig-Maximilians
 Universit\"at M\"unchen, Scheinerstr. 1, D-81679 M\"unchen,
 Germany\\ 
$^2$ Excellence Cluster Universe, Technical University Munich,
 Boltzmannstr. 2, D-85748 Garching, Germany\\ 
$^3$ Max-Planck-Institut f\"ur Extraterrestrische Physik,
 Giessenbachstr., D-85748 Garching, Germany\\ 
$^4$ Max-Planck-Institut f\"ur Astrophysik, Karl-Schwarzschild-Str. 1, D-85748 Garching, Germany\\ 
$^5$ Tartu Observatory, EST-61602 T\~oravere, Estonia\\
$^6$ Dipartimento di Fisica, Sezione di Astronomia, Universit\`{a} di Trieste, Via Tiepolo 11, I-34143 Trieste, Italy\\
$^7$ INAF-Osservatorio Astronomico di Trieste, Via Tiepolo 11, I-34143 Trieste, Italy}

\begin{document}

\date{Accepted 2013 June 12. Received 2013 May 31; in original form 2013 March 7}

\volume{434} \pagerange{684--695} \pubyear{2013}

\maketitle

\label{firstpage}

\begin{abstract}
We present the clustering of galaxy clusters as a useful addition to the common set of cosmological observables.
The clustering of clusters probes the large-scale structure of the Universe, extending galaxy clustering analysis to the high-peak, high-bias regime. Clustering of galaxy clusters complements the traditional cluster number counts and observable--mass relation analyses, significantly improving their constraining power by breaking existing calibration degeneracies.
We use the maxBCG galaxy clusters catalogue to constrain cosmological parameters and cross-calibrate the mass--observable relation, using cluster abundances in richness bins and  weak-lensing mass estimates.
 We then add the redshift-space power spectrum of the sample,  including an effective modelling of the weakly non-linear contribution and allowing for an arbitrary photometric redshift smoothing. The inclusion of the power spectrum data allows for an improved self-calibration of the scaling relation. 
We find that the inclusion of the power spectrum typically brings a $\sim 50$ per cent improvement in the errors on the fluctuation amplitude $\sigma_8$ and the matter density $\omegam$. Finally, we apply this method to constrain models of the early universe through the amount of primordial non-Gaussianity of the local type, using both the variation in the halo mass function and the variation in the cluster bias. We find a constraint on the amount of skewness $\fnl = 12 \pm 157 $ ($1\sigma$) from the cluster data alone.
\end{abstract}



\section{Introduction}
\label{i}

Galaxy clusters  are the most massive bound systems in the Universe which trace the evolution of the large-scale structure \citep[LSS; see the recent review by][]{Allen2011}. The initial density perturbations are thought to have formed in the early universe from inflationary physics \citep{Lyth2009}. In the simplest scenario, the perturbations can be modelled as Gaussian random fields \citep{BBKS1986}, which evolve gravitationally. This leads to the formation of bound dark matter structures -- the haloes, whose abundance is described by the halo mass function. The simplest infall formation model \citep{PS1974} is complicated by dynamical effects, meaning that accurate modelling of the mass function requires partial calibration \citep{ST1999,Maggiore2010,Corasaniti2011} or full fitting \citep{Jenkins2001,Tinker2008} to  $N$-body simulations.
Halo mergers and internal processes such as galaxy formation further complicate the picture at small scales \cite[e.g., see][for a review]{Borgani2009}.
Nonetheless, galaxy clusters form at a comoving scale of $\sim 10 \, h^{-1}$Mpc, allowing for a simpler theoretical description than is possible for smaller structures such as galaxies. Due to their scale, clusters reside in the tail of the halo mass function and thus their numbers are exponentially sensitive to variations in cosmology \citep[see e.g.,][]{Evrard1989,Frenk1990,Bahcall1997}.

Clusters are detected across multiple wavelengths with varying degrees of success. A few dozens have been found in the millimetre by the Atacama Cosmology Telescope \citep[ACT; ][]{ACT2012}, hundreds with the South Pole Telescope \citep[SPT; ][]{SPT2012} and the \emph{Planck} satellite \citep{Planck2011a}; also a few hundreds of clusters have been found in the X-ray \citep[REFLEX, BCS, eBCS catalogues by][]{Ebeling1998,Ebeling2000,Boeringer2004} using the \emph{ROSAT} satellite All-Sky Survey \citep[][]{Voges1999}, the \emph{Chandra} Cluster Cosmology Project \citep[][]{Burenin2007,Vikhlinin2009a} and by \citet{XMM2012} using \emph{X-ray Multi-Mirror Mission--Newton} \citep[][]{Fassbender2008}. Many tens of thousands have been found in the optical using Sloan Digital Sky Survey (SDSS) data to construct the maxBCG \citep{Koester2007a} and GMBCG \citep{Hao2010} catalogues, based on the selection of brightest cluster galaxies (BCGs) to identify the clusters' centres, and by \citet[][]{Gladders2005} using the Canada--France--Hawaii Telescope and Cerro Tololo Inter-American Observatory.

While detectable in large numbers, the main obstacle with using optical clusters as probes of cosmology is the difficult choice of a low-scatter mass proxy. 
Efforts on this front have been led by \cite{Rozo2010} and \cite{Zu2012}, who derived cosmological constraints from the maxBCG cluster sample. The tightest scaling relation between observable and cluster mass comes from X-ray data \citep[$<$10 per cent scatter;][]{Allen2008}. Constraints on dark energy with $\sim$20 per cent uncertainty were obtained from X-ray cluster samples studied by \cite{Mantz2008, Mantz2010} and \cite{Vikhlinin2009b}. 
Data on the cluster masses obtained from weak-lensing (WL) analyses of background galaxies have also been combined with the number counts to improve the constraining power of the cluster mass function \citep{Johnston2007,Sheldon2007,Mahdavi2007}. 
The statistics of rare events in the high-peak, high-mass limit has also been used by \citet{Hotchkiss2011} and \citet{Hoyle2012} to test cosmology. 

The uncertainty in the scaling relation is one of the biggest obstacles in using galaxy clusters as cosmological probes, as pointed out by \citet{Haiman2001} and \citet{Battye2003}. \citet{Majumdar2003} suggested to use the clustering of clusters as a complementary probe.
So far, only limited efforts have been dedicated to the measurement of the clustering properties of galaxy clusters: \citet{Huetsi2010} measured the power spectrum of maxBCG clusters resulting in weak detection of baryon acoustic oscillations (BAOs), \citet{Estrada2009} measured the correlation function for the same and \citet{Hong2012} measured the correlation function of the cluster catalogue by \citet{Wen2009}. Finally, \citet{Collins2000} measured the spatial correlation function of the REFLEX X-ray cluster catalogue, while \citet{Balaguera2011} measured its power spectrum. From the same survey, \citet{Schuecker2003} derived cosmological constraints from cluster abundances and large-scale clustering. The goal of this paper is to fully include the clustering information in the cosmological analysis of optical cluster data: we show that its inclusion significantly improves the cosmological constraints and also reduces the degeneracies between the scaling relation nuisance parameters. We present the improved cosmological results obtained in this way from the maxBCG data. 

As an interesting application, we present the constraining power of these data on the amount of primordial non-Gaussianity (PNG) of the initial density perturbations, which is expected to be produced in some models of the early universe. 
Briefly, while the simplest single-field slow-roll inflation produces nearly-Gaussian initial conditions \citep{Acquaviva2003,Maldacena2003}, there exist alternatives, such as multi-field models, which can produce large non-Gaussianities \citep[see e.g. the recent review by][]{Chen2010}. These would have multiple observable consequences, of which we here consider two: the halo mass function changes as a function of the non-zero skewness \citep{Matarrese2000,Loverde2008,Pillepich2010,Achitouv2012a,Achitouv2012b}, and in the local and orthogonal cases the halo bias becomes strongly scale-dependent due to the coupling of long- and short-wavelength modes \citep{Afshordi2008,Dalal2008,Matarrese2008,Slosar2008,Desjacques2009,Desjacques2010,Giannantonio2010,Schmidt2010,Desjacques2011}.
Measurements of PNG can potentially rule out entire classes of inflationary models \citep{Bartolo2004,Suyama2010}.
The latest constraints on the local PNG parameter $\fnl$ from the bispectrum of the cosmic microwave background (CMB) as measured by the \emph{Wilkinson Microwave Anisotropy Probe} (WMAP) satellite  are  $-3 < \fnl < 77$ at 95 per cent confidence level \citep{Hinshaw2012,Bennett2012}; comparable bounds have been obtained from the LSS using multiple galaxy catalogues \citep{Afshordi2008,Slosar2008,Xia2010a,Xia2010b,Xia2011a,Sefusatti2012,Ross2013,Giannantonio2013}; future galaxy surveys such as \emph{Euclid} are expected to reach an accuracy of $\Delta \fnl \sim 3 $ \citep{Giannantonio2012}.
\citet{Oguri2009} suggested that measuring the variance of cluster counts can yield significant constraints on PNG, while \citet{Sartoris2010} showed in principle how such constraints can be improved by using the cluster power spectrum. Forecasts for cosmology and PNG have been also investigated by \citet{Pillepich2012}, with the future \emph{eROSITA} X-ray cluster survey.
%

The paper is organized as follows. In Section \ref{sec:data}, we describe measurements of the cluster abundances, WL mass estimates and power spectrum of the maxBCG catalogue. In Section \ref{sec:model}, we introduce the theoretical framework including number counts and total mass determination from the mass function, mass--observable relation, bias and power spectrum definitions, and the effects of non-Gaussian initial conditions. Section \ref{sec:cosmomc} presents our Monte Carlo Markov Chain (MCMC) analysis, the cosmological constraints and the relevant degeneracies. We draw our conclusions in Section \ref{sec:concl}. We assume that the Universe is spatially flat on large scales throughout.

\section{Data}
\label{sec:data}

\subsection{The maxBCG cluster catalogue}
\label{sec:max}
\label{sec:maxcounts} 

The maxBCG catalogue \citep{Koester2007b} is a sample of 13,823 galaxy
clusters compiled from SDSS photometric
data. The catalogue is assembled by selecting the BCG and applying a red-sequence method to identify cluster
members in its neighbourhood. In this way clusters with richness
(number of member 
galaxies) ranging from 10 to 190 are selected. The low-mass limit of
this sample is $M_{\rm lim}\sim 7 \times 10^{13} h^{-1}
\mbox{M}_{\bigodot}$, which evolves weakly with redshift. This relatively
low-mass limit results in a sample that is significantly larger than
other current galaxy cluster catalogues. The clusters are chosen in an
approximately volume-limited way 
from a 500 $\mbox{Mpc}^{3}$ region, covering $\sim$7500 deg$^2$
of sky with a photometric redshift (photo-$z$) range of $0.1\leq z \leq0.3$. The
photo-$z$ errors are small and of the order of
$\Delta z=0.01$. An analysis of mock samples shows that the maxBCG
algorithm results in more than 90 per cent purity and more than 85 per cent
completeness, for clusters with masses $M\geq 10^{14}
\mbox{M}_{\bigodot}$ \citep{Koester2007b}.

We define the richness  $N_{\gal}$ as the number of red
galaxies within the radius $ R_{200}$ from the cluster
centre. $R_{200}$ is the radius within which the average overdensity
is 200 times the mean density of the Universe. 
The catalogue is divided into nine richness bins in the range of $11\leq N_{\gal}\leq
120$, which approximately corresponds to $7 \times 10^{13} \leq M \leq 1.2 \times 10^{15} h^{-1}
\mbox{M}_{\bigodot}$ \citep{Rozo2010}.  We found that adding the five
remaining high-mass clusters of the maxBCG with richness $N_{\gal} > 120$ 
has a negligible impact on the cosmological analysis, so we do not
include them. We also use an additional bin at $9 \leq N_{\gal}\leq 11$ (Rozo, private communication), although we checked that the results are not affected by this.
For the cosmological analysis, we include Poisson errors and sample
variance due to LSS \citep{Hu2003}.
Furthermore, we assume 100 per cent purity and completeness, including a 5 per cent uncertainty \citep{Rozo2010}, which we add in quadrature.
We found that the photo-$z$ errors have a negligible impact on the number counts analysis presented here, so we neglect their effect on the number counts covariance matrix.

\begin{figure}
\begin{center}
\includegraphics[width=\linewidth]{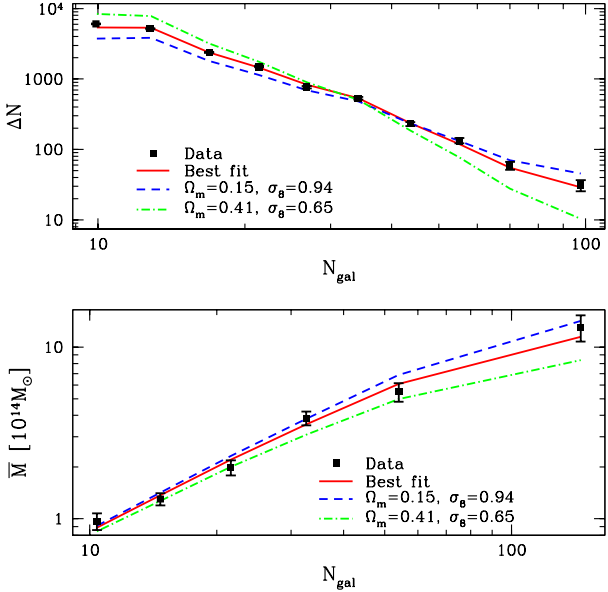}
\caption{Top panel: MaxBCG cluster counts data (black points) and theoretical predictions according to the prescriptions of Section~\ref{sec:model} for a choice of different cosmologies (without primordial non-Gaussianity). The red line represents the best-fitting model to our full data set (counts, total masses and power spectrum). Bottom panel: mean masses of maxBCG clusters (black points) and theoretical predictions for different cosmologies, as above.}
\label{fig:counts_extreme}
\end{center}
\end{figure}

The top panel of Fig.~\ref{fig:counts_extreme} shows the counts data, together
with the predicted counts for a selection of different
cosmologies, modelled as described in Section~\ref{sec:model}.

\subsection{Galaxy cluster masses from weak-lensing observations}
\label{sec:WLmass}

\citet{Sheldon2007}  measured the WL effect from
clusters in the maxBCG catalogue. By stacking the clusters, mean
cluster surface density profiles were created for different luminosity
and richness bins. The stacking of clusters in a given richness bin
improves the signal-to-noise ratio considerably compared to the measurement
of the profile of an individual cluster. \citet{Johnston2007} used these profiles and
reconstructed mean three-dimensional (3D) cluster density and mass profiles, which allows
one to estimate the mass (and concentration) of clusters in a given
redshift bin. For this reconstruction, a Navarro--Frenk--White (NFW)
profile \citep{NFW1997} for the cluster density was assumed. 
\citet{Johnston2007} were then able to construct a mean mass--richness
relation, finding for the whole sample of groups and clusters
\be
M_{200}(N_{\rm gal}) \simeq 8.8 \times 10^{13}h^{-1} \mbox{M}_{\bigodot} \, (N_{\rm gal}/20)^{1.28} \, ,
\ee
where $M_{200}$ is the mass contained within the radius $R_{200}$.
Due to photometric redshift bias, these masses are corrected upwards by
a factor of 1.18 as described in \cite{Mandelbaum2008a} and \cite{Rozo2010}.
For the cosmological analysis presented here, we follow
\citet{Rozo2010} and fit simultaneously for the mass--richness relation
using the \citet{Johnston2007} data and their errors. We use five richness bins
for this, in the range of $12\leq N_{\gal} \leq 300$, plus another extra bin at $9\leq N_{\gal} \leq 12$ (Rozo, private communication).

An independent WL analysis of the maxBCG sample was
performed by \citet{Mandelbaum2008b}, who found a mean mass difference of approximately $6$ per cent with respect to \citet{Johnston2007}. We follow \citet{Rozo2010} and include this discrepancy by introducing an offset factor $\beta$ with a suitable chosen prior, as described in equation~(\ref{eq:deltaNM}).

The bottom panel of Fig.~\ref{fig:counts_extreme} shows the mean WL mass estimates data,
together with the theoretical mean masses for a selection of different
cosmologies, modelled as described in Section~\ref{sec:model}. In general, the estimated WL mass of a galaxy cluster depends on the underlying cosmological model. To first order, for the analysis presented in this paper, this dependency is through the angular diameter distance, which is modified by the total matter density $\omegam$. In order to estimate the size of this cosmology dependence, we placed a galaxy cluster with mass $M = 1.1 \times 10^{15} h^{-1}\mbox{M}_{\bigodot}$ at redshift $z=0.2$ and produced a mock catalogue of sheared background galaxies. From this catalogue, we estimated the mass of the cluster by fitting to an NFW profile. We found that, if we allow $\omegam$ to change within the 1$\sigma$ level of our best-fitting cosmology, the mass varies within $5$ per cent. However, we allow for an uncertainty in the mass estimation with the offset factor $\beta$ with a prior width of $6$ per cent. Hence, any change due to a different $\omegam$ is completely degenerate with the $\beta$ parameter, which we conservatively assume does not depend on cosmology.


\subsection{MaxBCG power spectrum}
\label{sec:maxps}
We consider the redshift-space power spectrum of the maxBCG sample, as measured by \citet{Huetsi2010}. For the full details of the power spectrum measurement, along with systematics tests, we refer the reader to \citet{Huetsi2006b,Huetsi2006a,Huetsi2010}. 
The direct Fourier method by \citet{Feldman1994} (hereafter FKP) was used, with the difference that fast Fourier transforms (FFTs) were used instead of direct summation. This method actually yields the pseudo-spectrum, i.e. the measurement products are convolved with the window function of the survey. We take this into account when modelling the theoretical spectra in our analysis. To implement the modified FKP method, the following steps were followed.
\begin{enumerate}
\item The survey selection function (footprint) was represented using a random (unclustered) catalogue with 100 times more points than maxBCG sample.
\item The overdensity field was calculated on a regular grid using the triangular-shaped cloud method \citep{H&E1988} mass assignment scheme -- it was checked that the aliasing effects due to the finite grid size were negligible for the measurements and were nonetheless corrected with the iterative method by \citet{Jing2005}.
\item The gridded overdensity field was transformed into Fourier space using the FFT.
\item The raw 3D power spectrum was estimated by taking the modulus squared of the FFT.
\item The shot-noise contribution was subtracted.
\end{enumerate}

The uncertainties on the power spectrum measurements were estimated with three different methods: the original FKP theoretical prescription, which assumes Gaussian errors from cosmic variance plus a shot-noise contribution; a jackknife method, implemented by dividing the survey into a total of 75 regions; and a Monte Carlo (MC) method, based on the fiducial $\Lambda$ cold dark matter ($\Lambda$CDM) cosmology, in which 1000 mock realizations of the maxBCG survey were generated, including redshift-space distortions (RSDs) and photo-$z$ errors.
These three methods were shown to be comparable; in this work we use the MC covariance matrix. 

\begin{figure}
\begin{center}
\includegraphics[width=\linewidth]{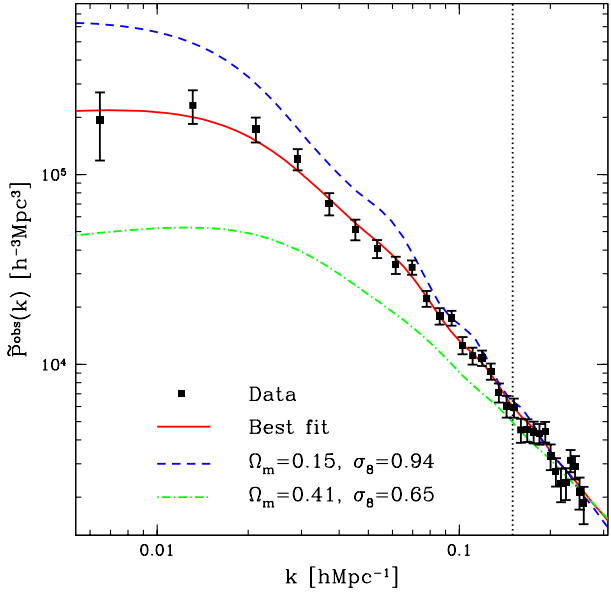}
\caption{The observed power spectrum of maxBGC clusters (black points) compared with the full theoretical modelling $\widetilde{P}^{\mathrm{obs}}$ described in Section~\ref{sec:model} for our best-fitting model (red solid line), and for two other models, assuming no primordial non-Gaussianity. The dotted line at $k=0.15 \, h$ Mpc$^{-1}$ represents our choice of $k_{\max}$, which is the smallest scale we use in the analysis.}
\label{fig:pkfull_extreme}
\end{center}
\end{figure}

The power spectrum measurements are shown in Fig.~\ref{fig:pkfull_extreme}. To take into account data in the quasi-linear regime only, we restrict ourselves to scales larger than (wavenumbers smaller than) $k_{\max} = 0.15 \, h\,$Mpc$^{-1}$.

\subsection{The off-diagonal covariance}
\label{sec:err}

The diagonal blocks of the data covariance matrix for counts, masses and the power spectrum have been described above.
We estimate the off-diagonal terms of the covariance matrix between the clustering and the binned number distributions $\Delta N$ of the maxBCG clusters using jackknifes. For simplicity,  instead of the power spectrum we use here,  as a clustering estimator, the projected correlation function $w(\theta)$ defined as
\be
w(\theta) \equiv \langle \delta_h( \hat{\vec{n}} ) \, \delta_h( \hat{\vec{n'}}) \rangle \, ,
\ee
where $\delta_h( \hat{\vec{n}} )$ is the halo (cluster) projected overdensity in a direction $\hat{\vec{n}}$, and the average is carried over all pairs at an angular distance $\theta$. We select $\theta$ within a range representative of the $k$ scales of interest at $z\sim0.2$.

We use the jackknife technique as follows: we split the maxBCG footprint into 100 equal-area jackknife regions using \textsc{healpix}\footnote{http://healpix.jpl.nasa.gov/}\citep[]{Gorski2005} and populate the full footprint with 50 random points for each maxBCG cluster, to reduce shot noise. We use the  correlation estimator by \citet{Landy1993} to calculate the correlation function, and bin the number of clusters within six equal-width bins in $\log_{10}$ space.  We iteratively remove and replace each jackknife region and calculate the number histogram and correlation function at each iteration. 
The covariance matrix $C_{\mathrm{JK}}$ between measured statistics $x=x(\alpha)$ and $y=y(\beta)$ can be estimated from $N$ jackknifes using \citep[see e.g.][]{Efron1982}
\begin{equation}
\left[ C_{\mathrm{JK}}(x_i,y_j) \right]_{\alpha,\beta} = \frac{N-1}{N}\sum_{k=1}^{N}( x^k_{-i} -\bar x_{i}  )_{\alpha} \, ( y^k_{-j} - \bar y_{j}  )_{\beta} \, ,
\end{equation}
where $x_{-i}$ ($y_{-j}$) is the statistic with jackknife region $i$ $(j)$ removed and $\bar{x}_{i}$ ($\bar{y}_{j}$) is the average value of all $x_{-i}$ ($y_{-i}$). We note that typically, but not necessarily, $x$ and $y$ are the same statistic.

\begin{figure}
   \begin{center}
   \includegraphics[width=\linewidth]{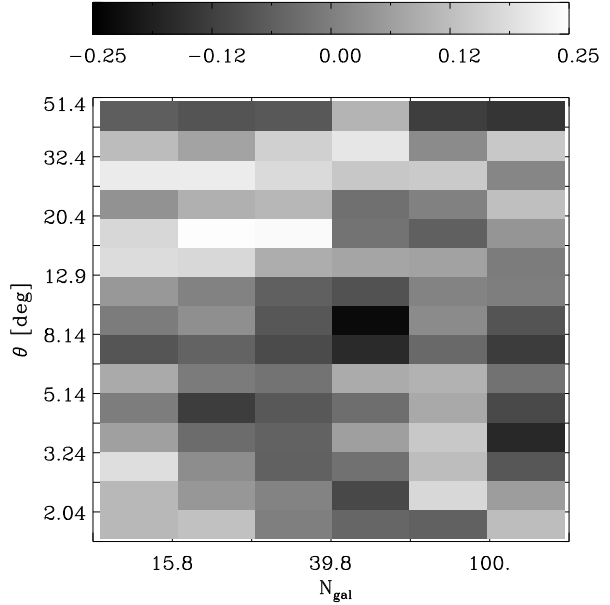}  
   \caption{The off-diagonal elements of the normalized covariance matrix of the correlation function $w(\theta)$ and the histogram distribution of $\Delta N$, as calculated using the jackknife technique.}
   \label{offdiag}  
   \end{center}
\end{figure}

We compare the square root of the diagonal elements of the  covariance matrix $C_{\mathrm{JK}}\big[w(\theta), w(\theta) \big]$ with the error expected from Poisson counting statistics and find agreement with the theoretical expectations \citep[as described by e.g.,][]{Ross2009}, 
and also find that the diagonal elements of $C_{\mathrm{JK}}\big(\Delta N, \Delta N \big)$ are approximately Poissonian, independently of the number of jackknifes used. In Fig.~\ref{offdiag}, we show the off-diagonal terms of the normalized $C_{\mathrm{JK}}\big[\Delta N, w(\theta) \big]$, and note that the average value and 1$\sigma$ error of the off-diagonal terms are $-0.03 \pm 0.10$, which is consistent with zero. We observe that as the number of jackknifes increases, the mean and error of the average value of the off-diagonal terms approaches and fluctuates around zero.

We compare the magnitude of the off-diagonal terms obtained from the maxBCG clusters with simulated clusters from the Millennium Simulation \citep{Springel2005,Lemson2006}. Specifically, we join the light-cone table of {\tt Henriques2012a.wmap1.BC03\_AllSky\_00} \citep{Guo2011,Henriques2012} with the halo-tree table {\tt MPAHaloTrees..MHalo}. We apply the same redshift and survey footprint constraints to mimic the maxBCG sample and calculate the correlation function and histogram distribution of $\Delta N$. We find the data and simulations to agree closely: e.g., for 100 jackknifes the mean and 1$\sigma$ error of the off-diagonal terms are $0.00 \pm 0.10$ from the simulations.

From these tests, we conclude that ignoring the off-diagonal covariance matrix between clustering and number counts is a reasonable approximation.

\section{Theoretical modelling of cluster statistics} 
\label{sec:model}

\subsection{The cluster mass function}
\label{sec:mass function}

\citet{PS1974} first calculated the expected number of dark matter haloes of a given mass and redshift. This was better described by the excursion set approach \citep{Bond1991} and generalized to non-spherical model by \citet{ST1999}, who calibrated their mass function with $N$-body simulations. Even more accurate estimations are achieved with a full fitting to $N$-body simulations \citep[i.e.][]{Jenkins2001}. The current state-of-the-art halo mass function has been estimated by \citet{Tinker2008,Tinker2010}: this mass function is valid over wide redshift and mass ranges.

The expected number density of virialized dark matter haloes as a
function of mass and redshift can be expressed as
\begin{equation}
\frac{\mathrm{d}n(M,z)}{\mathrm{d}\ln M}=\bar{\rho}_{\mathrm{m}} \left|\frac{\mathrm{d}\ln
\sigma^{-1}}{\mathrm{d}M}\right| f(\nu)\, ,
\label{eq:massfn}
\end{equation}
where $\bar{\rho}_{\mathrm{m}}$ is the mean matter density of the Universe, $\nu \equiv \delta_c/\sigma(M,z)$, $\delta_c=1.686$ is the threshold linear overdensity for spherical collapse in a matter-dominated Universe and $\sigma^2(M,z)$ is the variance of the linear matter density field at $M = 4 \pi R^3 \bar{\rho}_m / 3$. 
In this work we use the mass function given by \citet{Tinker2010} for cluster mass at $R_{200}$, with an overdensity of $\Delta=200$ in units of the mean mass density of the Universe:
\begin{equation}
f_T(\nu) = 0.368 \left[1+\left(\hat{\beta}\nu\right)^{-2\hat{\phi}}\right]\nu^{2\hat{\eta}+1} {\mathrm{e}}^{-\hat{\gamma}\nu^2/2}\ ,
\label{eq:fT}
\end{equation}
where the parameters evolve in redshift as
\[
\hat{\beta}=0.589 \, (1+z)^{0.20} \, , \:\:\:\:\: \hat{\phi}=-0.729 \, (1+z)^{-0.08} \, , 
\]
\begin{equation}
\label{eq:tinkerparams}
\hat{\eta}=-0.243 \, (1+z)^{0.27} \, , \:\:\:\:\:  \hat{\gamma}=0.864 \, (1+z)^{-0.01}\, .
\end{equation}

\subsection{The mass--richness scaling relation}
\label{sec:scal}
In order to perform a cosmological analysis, we need to make some
assumptions on the scaling relation between the true mass $M$ of a cluster and its richness $N_{\rm gal}$.
We first consider the probability of observing $N^{\obs}_{\gal}$ member galaxies at $R_{200}$ for a given true mass $M$ of the cluster. This can be written as 
\begin{equation}
\label{eq:pNobsM}
p(N^{\obs}_{\gal}|M) = \int p(N^{\obs}_{\gal}|N_{\gal}) \, p(N_{\gal}|M) \, \mathrm{d}N_{\gal} \, ,
\end{equation}
where $p(N_{\gal}|M)$ is a delta function, because the relation between $M$ and $N_{\gal}$ is given as follows. Following \citet{Johnston2007} and \citet{Rozo2010}, we assume the scaling relation to be a power law in mass, i.e. 
\begin{equation}
\label{eq:scalingRelation}
\ln M\ =\ \ln M_{200|20} +\alpha_{N}\ln (N_{\gal}/20)\ \ ,
\end{equation}
with $M_{200|20}$ the mass of a cluster with 20 member galaxies within a radius of $R_{200}$ and $\alpha_N$ the slope of the relation. This provides the mean of the distribution between $N_{\gal}^{\obs}$ and $M$. We fit this relation by fixing two pivot points in mass $M_1 = 1.3 \times 10^{14}$ and $M_2 = 1.3 \times 10^{15} \, \mbox{M}_{\bigodot}$, while
the corresponding richness values $\ln N_1 \equiv \ln N_{\gal}|M_1$ and $\ln N_2 \equiv \ln N_{\gal}|M_2$ are kept as free parameters.

We then assume $p(N^{\obs}_{\gal}|N_{\gal})$ to follow a lognormal distribution as suggested by \citet{Lima2005} 
\begin{equation}
\label{eq:pNobsN} 
p(N^{\obs}_{\gal}|N_{\gal}) = \frac{1}{\sqrt{2\pi \sigma^2_{\ln N^{\obs}_{\gal}| M } }}\ \exp\left[-x^2(N^{\obs}_{\gal})\right]\ ,
\end{equation}
where 
\begin{equation}
\label{eq:x}
x(N^{\obs}_{\gal}) = \frac{\ln N^{\obs}_{\gal}-\ln N_{\gal}(M)}{\sqrt{2\sigma^2_{\ln N^{\obs}_{\gal}| M }  }} 
\end{equation}
and $\sigma_{\ln N^{\obs}_{\gal}| M }$ is the scatter around the mean $N_{\gal}(M)$ given by equation~(\ref{eq:scalingRelation}) \citep{Battye2003,Lima2005}. Note that we used $\sigma_{\ln N^{\obs}_{\gal}|N_{\gal} }=\sigma_{\ln N^{\obs}_{\gal}| M}$, which holds because $p(N_{\gal}|M)$ is a delta function. The statistical scatter around the scaling relation is assumed to be constant with redshift and mass for individual clusters. To obtain an estimate of this quantity is not trivial; however, \citet{Rozo2009} used WL and X-ray observations together with the maxBCG richness to have three different mass proxies. By demanding consistency between the X-ray and WL measurements, \citet{Rozo2009} found $\sigma_{\ln M | N^{\obs}_{\gal}}=0.45^{+0.20}_{-0.18}$, which is the scatter in mass given the richness. For our cosmological analysis described in Section~\ref{sec:cosmomc}, we need to place a prior on the converse scatter, $\sigma_{\ln N^{\obs}_{\gal}|M}$. The two quantities can be readily related to each other by invoking the relation of equation~(\ref{eq:scalingRelation}), which results in $\sigma_{\ln M| N^{\obs}_{\gal}} = \alpha_{N} \, \sigma_{\ln N^{\obs}_{\gal} | M}$.

\subsection{Modelling galaxy cluster counts and total masses}
\label{sec:counts}
In order to predict the number of observed galaxy clusters for an
observed richness $N_{\gal}^{\obs}$, we can use the probability
distribution and scaling relation defined in the previous section. 
The cluster average number density within a richness bin $[N^{\obs}_{\gal, \, i} \, , \, \ N^{\obs}_{\gal,\, i+1}]$ is given by
\begin{eqnarray}
n_i &=& \int^{N^{\obs}_{\gal,\, i+1}}_{N^{\obs}_{\gal,\, i}} \mathrm{d}\ln N^{\obs}_{\gal} \int \mathrm{d}\ln N_{\gal} \frac{\mathrm{d}n}{\mathrm{d}\ln N_{\gal}} p(N^{\obs}_{\gal}| N_{\gal}) = \nonumber \\ 
&=& \int \mathrm{d}\ln N_{\gal}\ \frac{\mathrm{d}n}{\mathrm{d}\ln N_{\gal}}\ \frac{1}{2}\ [\mbox{erfc}(x_i) - \mbox{erfc}(x_{i+1})]\ ,
\label{eq:ni}
\end{eqnarray}
where $x_i \equiv x(N^{\obs}_{\gal, \, i})$, 
\be
\frac{\mathrm{d}n}{\mathrm{d}\ln N_{\gal}} = \frac{\mathrm{d}n}{\mathrm{d}\ln M} \frac{\mathrm{d}\ln M} {\mathrm{d}\ln N_{\gal}} = \alpha_N \, \frac{\mathrm{d}n}{\mathrm{d}\ln M} \, ,
\ee 
and we have employed the scaling relation of equation~(\ref{eq:scalingRelation}).
The total number of predicted galaxy clusters within a
richness bin can be calculated as
\begin{equation}
\label{eq:deltaN}
\Delta N_i = \Delta\Omega\int_{z_{\min}}^{z_{\max}} \mathrm{d}z \,  \frac{\mathrm{d}^2V}{\mathrm{d}z \, \mathrm{d}\Omega} n_i\ ,
\end{equation}
where $\Delta\Omega$ is the survey sky coverage and $\mathrm{d}^2V/\mathrm{d}z/\mathrm{d}\Omega$ is the volume element. The cosmology dependence is driven by the
mass function and by the comoving volume element.

\label{sec:masses}

We can write similar expressions to equations~(\ref{eq:ni}) and (\ref{eq:deltaN}) for the total mass of clusters. The average total mass $(nm)_i$ contained  within the same richness bin can be obtained as in equation~(\ref{eq:ni}) by weighting the integrand by the mass, estimated via the mass--observable relation.
The total mass of clusters  within a richness bin is then
\begin{equation}
 \label{eq:deltaNM}
\left(\Delta N\bar{M} \right)_i = \beta\ \Delta\Omega\int_{z_{\min}}^{z_{\max}} \mathrm{d}z\ \frac{\mathrm{d}^2V}{\mathrm{d}z \, \mathrm{d}\Omega} (nm)_i \, ,
\end{equation}
where $\beta$ is an additional nuisance parameter introduced to account for possible mismatch with the WL masses, as discussed above and in \cite{Rozo2010}.

\subsection{Clustering of clusters}
\label{sec:PS}

Galaxy clusters can be studied as tracers of the LSS \citep{Mo1996a}, corresponding to the highest-density regions of the dark matter overdensity field $\delta(\vec{x},z)$. 
If we assume linear theory and a local deterministic halo bias \citep{Fry1993}, then the dark-matter haloes overdensity is $\delta_h(\vec{x},M,z) = b_0 + b_h (M,z) \, \delta(\vec{x},z)$. The local bias assumption breaks down in the case of PNG (see Section~\ref{sec:fnl}). As the effect of baryons is negligible
for the clustering properties of the clusters, in the following we use the naming `cluster' and `halo' interchangeably.

\subsubsection{Halo bias}

\begin{figure}
\begin{center} 
\includegraphics[trim = 0 275 0 0, clip, width=\linewidth]{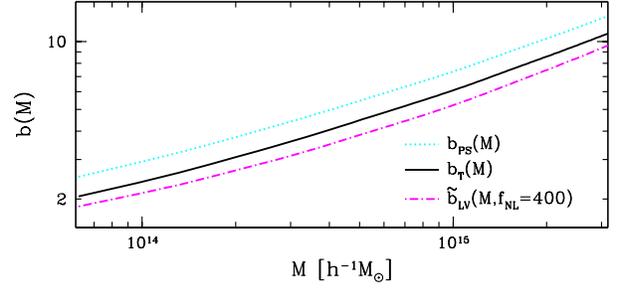}
\caption{Mass dependence of the  linear halo bias at $\bar z = 0.2$ for three mass functions: Press--Schechter (cyan dotted), Tinker (black solid) and modified LoVerde mass function in the presence of PNG (magenta dot--dashed), with $\fnl = 400$. Cosmology is fixed to our combined best-fitting model.}
\label{fig:pklin_extreme}
\label{fig:bias_extreme}
\end{center}
\end{figure}

The halo bias can be derived from a theory of the mass function via the peak-background split formalism \citep{Cole1989,Mo1996b}. This method gives a prediction for the halo bias in Lagrangian space $ b^L (M,z)$, which can be evolved into the observable Eulerian space as $b = 1 + b^L$ \citep{Mo1996b}, considering linear perturbation only, spherical collapse and no large-scale velocity bias. We assume that the bias is scale independent (except for the modifications in the presence of PNG).
At linear order, the Lagrangian bias is 
\be \label{eq:bfromn}
b^L(M,z) = - \frac{f'(\nu)}{f(\nu) \, \sigma} \, ,
\ee
where the derivative of the mass function is taken with respect to $\nu$, and the mass and redshift dependences of $\nu$ and $\sigma$ are implicit. 
When using the Tinker mass function and keeping the leading order terms, the Eulerian bias is \citep{Tinker2010}
\begin{equation}
 \label{eq:bT}
b_{T}(M,z) \simeq 1+ \frac{\hat{\gamma}\nu^{2} - (1+2\hat{\eta})}{\delta_{c}}+\frac{2\hat{\phi}/\delta_{c}}{1+[\hat{\beta}\nu]^{2\hat{\phi}}}  \, ,
\end{equation}
where all parameters are defined as in equation~(\ref{eq:tinkerparams}).
In Fig.~\ref{fig:bias_extreme}, we show the Tinker halo bias $b_T (M)$ as a function of halo mass $M$ at $\bar z=0.2$, which is the mean redshift of the maxBCG clusters, compared with the \citet{PS1974} (hereafter PS) case $   b_{\mathrm{PS}} = 1 - 1 / \delta_c  + \delta_c / \sigma^2 $ and with the scale-independent part of the bias in the presence of PNG described below, for our combined best-fitting model.

We obtain the average cluster bias $\bar b $ over the mass range of our data by weighting with the mass function \citep{Lima2005}:
\begin{equation}
\label{eq:bbar}
\bar b (z) = \frac{1}{\Delta N}  \, \int_{M_{\min}}^{M_{\max}} \mathrm{d}\ln M\ \frac{\mathrm{d}n(M, z)}{\mathrm{d}\ln M}  \, b_T(M, z) \ ,
\end{equation}
where the normalization factor $\Delta N$ is the full integral of the  mass function in the observed range.

\subsubsection{Power spectra}

We then define the observable clustering statistics in Fourier space (denoted by a tilde). As we consider linear scales only, the halo--halo power spectrum $P_{\mathrm{hh}}$ can be related to the linear matter power spectrum $P_{\mathrm{lin}}$ as
\be
\label{eq:pM}
 P_{\mathrm{hh}}(k, M, z) =  b^2 (M,z) \, P_{\mathrm{lin}}(k, z) =  b^2 (M,z) \, D^2(z) \, P_{\mathrm{lin}}(k, 0) \, ,
\ee
where $D(z)$ is the linear growth function. We integrate the mass dependence by weighting the bias as described in equation~(\ref{eq:bbar}) and we compute all quantities at the mean redshift of our cluster sample, $\bar z \simeq 0.2$. This is further justified by observing that the growth of $\bar b(z)$ is compensated by a similar drop in $D(z)$; we have checked that for our fiducial cosmology, in the observed range $0.1 \le z \le 0.3$, the variation of $\bar b(z) \, D(z)$ is at the per cent level. 
 
Before fitting models to the data, the following four effects have to be taken into account, following the description by \cite{Huetsi2010}: the photo-$z$ errors, which are responsible for a damping of the spectrum on small scales; the convolution with the survey window, which suppresses the power on large scales;  the non-linearities which add power on small scales; and the RSDs.
The total observed power spectrum $P^{\mathrm{obs}}$ is modelled as
\begin{equation}
{P^{\mathrm{obs}}}(k)=\int \mathrm{d}\ln \kappa \, \kappa^{3} \, P_{\mathrm{NL}}(\kappa) \, K(\kappa,k) \, ,
\label{eq:pfinal}
\end{equation}
where $K(\kappa,k)$ is the kernel accounting for the effect of the finite survey area, given in equations~(9)--(11) of \cite{Huetsi2010}, and $P_{\mathrm{NL}}$ contains the remaining corrections and the effect of non-linearities. In our analysis, we only use data up to $k_{\max} = 0.15 \, h $ Mpc$^{-1}$ and we follow  \cite{Huetsi2010}, modelling the effect of residual weak non-linearities with a simple effective fitting function with one free parameter $\qnl$.
All these contributions lead to
\be
\label{eq:pnl}
 P_{\mathrm{NL}}(k)=\left(b^{\mathrm{obs}}\right)^2(1+q_{\mathrm{NL}}k^{3/2}) \ s(k) \ P_{\mathrm{lin}}(k) \left[1+\frac{2}{3}\beta_z+\frac{1}{5}\beta_z^2\right] \, .
\ee
Here the bias is rescaled as $b^{\mathrm{obs}} = \bar{b}\,B$, where we include a nuisance parameter $B$  to represent the uncertainty on the bias derived from the mass function. We model the photo-$z$ smoothing with a corrective factor
\be
s(k) = \left( \frac{\sqrt{\pi}}  {2 \, \sigma_{z} \, k} \right) \,  \mathrm{erf} \, (\sigma_{z} \, k) \, ,
\ee
assuming that photo-$z$ errors follow a Gaussian distribution with dispersion $\delta z$  and corresponding spatial smoothing scale $\sigma_z = \delta z\,c/H_0$.                                                      
The last term of equation~(\ref{eq:pnl}) is the correction due to RSDs, for which we assume $\beta_z (\bar z) \simeq \omegam^{0.55} (\bar z) /b^{\mathrm{obs}} (\bar z) $ \citep{Kaiser1987}. We have checked that the RSD correction changes at most at the per cent level if we calculate it at the limits of our redshift range.
We finally take into account the Alcock--Paczynski effect \citep{AP1979}: we rescale the full theoretical power spectrum with respect to the cosmology used to convert redshifts to distances in the measurements (denoted by the superscript `fid'), assuming that a single isotropic dilation applies \citep{Eisenstein2005,Huetsi2006c}, i.e.
\be
\widetilde{P}^{\mathrm{obs}} (k) = \frac {1} {c^3_{\mathrm{isotr}}} \, P^{\mathrm{obs}} \left( \frac{k}{c_{\mathrm{isotr}}} \right) \, .
\ee
Here, $ c_{\mathrm{isotr}} = \left( c_{\parallel} \, c^2_{\perp} \right)^{1/3} $, $ c_{\parallel} = H^{\mathrm{fid}} / H$, $c_{\perp} = D_A / D_A^{\mathrm{fid}}$ and $D_A$ is the angular diameter distance, where all quantities are calculated at the mean redshift $\bar z$.

In Fig.~\ref{fig:pkfull_extreme}, we show the full cluster power spectrum ${\widetilde{P}^{\mathrm{obs}}}(k)$. The different lines correspond to the theory curves for our combined best-fitting cosmology (red solid) and for two other models (blue and green) chosen to be at the $2\sigma$ limit of the marginalized $\omegam-\sigma_8$ contour, compared with the data.

\subsection{Primordial non-Gaussianity} 
\label{sec:fnl}

We extend our model to constrain PNG from both bias and abundances of the maxBCG clusters.
Briefly, following e.g. the notation of \citet{Giannantonio2010}, we introduce the $\fnl$ parameter to quantify the amount of PNG in the simplest local, scale-independent case as
\begin{equation}
\label{eq:potential}
\Phi(\vec{x}, z_{\ast}) = \varphi(\vec{x}, z_{\ast}) + \fnl  \left[ \varphi^2(\vec{x}, z_{\ast}) -  \langle  \varphi^2  \rangle(z_{\ast})  \right] \, ,
\end{equation}
where $\Phi$ is the Bardeen's potential at a primordial redshift $z_{\ast}$ and $\varphi$ is an auxiliary Gaussian potential. Throughout this paper, we define $\fnl$ by writing the previous equation at early times (i.e. $z_{\ast}\approx1100$).
The potential power spectrum can be approximated by its Gaussian part, $P_{\Phi}(k) \simeq P_{\varphi}(k)$, at leading order in $\fnl$ and neglecting trispectrum corrections.
The matter perturbations are related to the primordial potential by the Poisson equation:
\be \label{eq:poisson}
\tilde \delta (\vec k, z) = \alpha(\vec k, z) \, \tilde \Phi(\vec k, z_{\ast}) \, ,
\ee
with
\be
\alpha(k,z) = \frac{2 \, c^2 \, k^2 \, T(k) \, D(z) \, g(0)}{3 \, \omegam \, H_0^2 \, g(z_{\ast})} \, .
\ee
Here, $T(k)$ is the transfer function and $g(z) \propto (1+z) \, D(z)$ is the growth function of the potential. We can then write for the matter power spectrum $P$
\be
P(k,z) = \alpha^2(k,z) \, P_{\Phi}(k, z_{\ast}) \simeq  \alpha^2(k,z) \, P_{\varphi}(k, z_{\ast}) \, ;
\ee
we  consider linear theory only, so we assume $P = P_{\mathrm{lin}}$.

The halo mass function is modified in the presence of PNG as it gains a dependence on the skewness. We use the \cite{Loverde2008} mass function (LV), which was obtained by using the Edgeworth expansion and is given by 
\begin{equation}
\label{eq:fLV}
f_{\mathrm{LV}}(\nu) = \sqrt{\frac{2}{\pi}} \mathrm{e}^{-\frac{\nu^2}{2}} \left[ \nu+S_3 \frac{\sigma}{6} (\nu^4-2\nu^2-1) + \frac{dS_3}{d\ln \sigma} \frac{\sigma}{6}(\nu^2-1)\right] \, ,
\end{equation}
where $S_3$ is the skewness of the matter density field defined as in \citet{Desjacques2009} (the mass dependence  is implicit).
To improve the agreement with $N$-body simulations, and for consistency with the rest of our analysis, we replace its Gaussian limit from the PS to the Tinker form, so that we use the rescaled form defined as
\begin{equation}
\label{eq:fLVtilde}
\widetilde f_{\mathrm{LV}}(\nu) \equiv \frac{f_{\mathrm{T}}(\nu)}{f_{\mathrm{PS}}(\nu)} \, f_{\mathrm{LV}}(\nu)\ , 
\end{equation}
where $f_{\mathrm{PS}} = \sqrt{2 /\pi} \, \nu \,  \exp\left(-\nu^2/2 \right)$  is the PS mass function.

We apply the peak-background split formalism and analytically derive the Lagrangian linear halo bias associated with the LoVerde mass function using equation~(\ref{eq:bfromn}) as 
\begin{equation}
\label{eq:bLV}
b^L_{\mathrm{LV}}(\nu) = \frac{\delta_c}{\sigma^2}-\frac{1}{\sigma} \frac{6 + S_3\sigma(4\nu^3-4\nu)+ 2 \frac{dS_3}{d\ln \sigma} \sigma \nu}{6\nu+S_3\sigma(\nu^4-2\nu^2-1)+ \frac{dS_3}{d\ln \sigma}\sigma(\nu^2-1)}\ ,
\end{equation}
while the Lagrangian bias associated with the rescaled mass function of equation~(\ref{eq:fLVtilde}) is given as
\be
\label{eq:bLVtilde}
\widetilde b^L_{\mathrm{LV}}(\nu) = b^L_{\mathrm{LV}}(\nu) + b^L_{\mathrm{T}}(\nu) - b^L_{\mathrm{PS}}(\nu) \, ,
\ee
which is the bias we use in the following.

In the presence of PNG, the halo density perturbations depend not only on the dark matter perturbations $\delta$, but also on the potential $\varphi$. The latter can then be related back to the density in Fourier space by using the Poisson equation, so that the effective Eulerian bias 
can be written at the mean redshift $\bar z \simeq 0.2$ as
\be
b_{\mathrm{eff}}(M,k,\fnl) = b(M,\fnl) + \Delta b (M,k, \fnl)  \, ,
\ee
where the bias contains implicitly a scale-independent correction $\delta b(\fnl) \equiv b(M,\fnl) - b(M,0) $ with respect to the Gaussian case, following from the difference in the mass function,
and the scale-dependent part is 
\begin{equation}
\label{eq:deltab2}
\Delta b(M,k,\fnl) = \frac{2 \, \fnl \, \delta_c \,  b^L(M,\fnl)} {\alpha(k, \bar z)} \, .
\end{equation}

As in the Gaussian case, we average the bias over the masses in our catalogue following equation~(\ref{eq:bbar}).
In order to take into account the uncertainty on our assumption of a mass function, we also introduce a nuisance parameter $B$ as in Section~\ref{sec:PS}, which rescales the bias as $b^{\mathrm{obs}} = \bar{b}\,B $.

The scale-independent correction $\delta b (\fnl)$ is small, easily confused with other normalization effects, and relies on the assumed form of the mass function and the peak-background split method. For these reasons, it is worth ensuring that the results do not depend on this contribution. We make sure this happens in our case because any constant rescaling of the bias can be equally explained by either a change in the nuisance parameter $B$ or a change in $\fnl$. But since a model with $\fnl \ne 0$ also predicts the scale-dependent bias, it will be favoured only in case such a feature is indeed observed in the data, otherwise the $B \ne 1$ model will be assigned a better likelihood.
In practice, we impose some Gaussian priors centred on $B = 1$, but we have checked that the results on $\fnl$ do not depend significantly on this choice.

\begin{figure}
\begin{center}
\includegraphics[width=\linewidth]{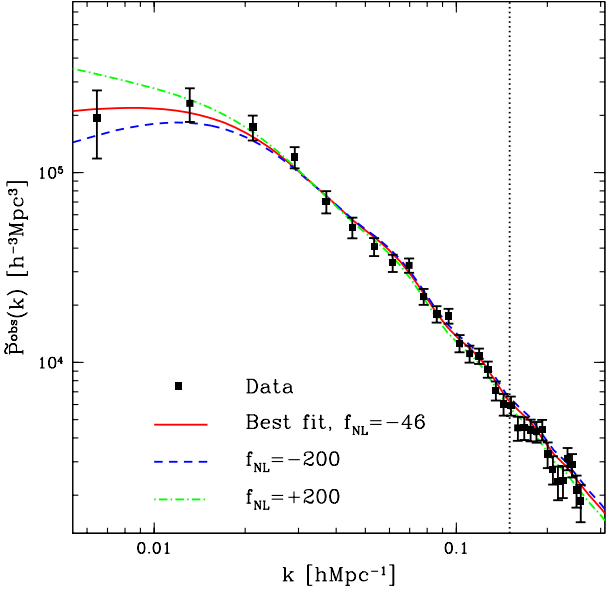}
\caption{The effect of PNG on the cluster power spectrum. We compare the data (black points) with the predictions for the best-fitting model to our data with $\fnl=-46$ (red solid) and for two cases with $\fnl=-200$ (blue dashed) and $\fnl=200$ (green dot--dashed). The dotted line at $k_{\max}=0.15 \, h$ Mpc$^{-1}$ represents the smallest scale we use in the analysis. }
\label{fig:pkfull_fnl_extreme}
\end{center}
\end{figure}

In Fig.~\ref{fig:pkfull_fnl_extreme}, we show the full power spectrum $\widetilde{P}^{\mathrm{obs}}(k)$ in the presence of PNG for a choice of $\fnl$ values, compared with the data. The scale-dependent bias induced by PNG is visible on large scales (small $k$), while the smaller scale-independent contribution can be seen on small scales (large $k$). Note that the survey window convolution of equation~(\ref{eq:pfinal}) partially suppresses the effect of PNG on the largest scales, which become comparable with the survey volume.

\section{Likelihood analysis and results}
\label{sec:cosmomc}

We use cluster counts, WL masses and the cluster power spectrum to fit the richness--mass relation and constrain cosmology simultaneously. In particular, our observables are given as follows.
\begin{enumerate}
	\item [(1)] Cluster counts $\Delta N$, divided into 10 richness bins.
	\item [(2)] Total mass of clusters $\Delta N \bar{M}$, divided into 6 richness bins.
	\item [(3)] Cluster power spectrum $\widetilde{P}^{\mathrm{obs}}$, divided into 18 $k$ bins.
\end{enumerate}
The covariance matrix we use for the cosmological analysis is composed by the parts discussed in Section \ref{sec:data}.
In addition to the cluster data, we also use the CMB power spectra from \emph{WMAP7} \citep{Larson:2011a}, in the cases specified below.

We assume a flat $\Lambda$CDM cosmological model. When using cluster data alone we fix the Hubble parameter $h=0.7$, primordial spectral index $n_{\mathrm{s}}=0.96$ and baryon density $\Omega_{\mathrm{b}}=0.044$, as these parameters are not easily constrained in this case; we relax these assumptions when adding external CMB data. Note that we need to fix the spectral index of scalar density perturbations because of the small range in scale which our mass range corresponds to. 

We then perform Bayesian parameter estimation by running MCMCs, using Metropolis sampling with a modified version of the \textsc{cosmomc} code \citep{LewisBridle2002}. In Table \ref{tab:Table4parameters}, we list all the parameters of the analysis, including their assumed priors.
We  estimate the posterior probability distributions in the following cases. 
\begin{enumerate}
 \item \textbf{Counts only}: six free parameters [$\omegacdm$, $\log(10^{10} A_{\mathrm{s}})$, $\ln N_1$, $\ln N_2$, $\sigma_{\ln M| N^{\obs}_{\gal}}$, $\beta$], without the cluster power spectrum.
 \item \textbf{Counts with \boldmath{$\fnl$}}: seven free parameters [$\omegacdm$, $\log(10^{10} A_{\mathrm{s}})$, $\ln N_1$, $\ln N_2$, $\sigma_{\ln M| N^{\obs}_{\gal}}$, $\beta$, $\fnl$], without the cluster power spectrum.
 \item \textbf{Counts+\boldmath{$P(k)$}}: nine free parameters [$\omegacdm$, $\log(10^{10} A_{\mathrm{s}})$, $\ln N_1$, $\ln N_2$, $\sigma_{\ln M| N^{\obs}_{\gal} }$, $\beta$, $\qnl$, $\sigma_{z}$, $B$], with  the cluster power spectrum.
 \item \textbf{Counts+\boldmath{$P(k)$} with \boldmath{$\fnl$}}: 10 free parameters [$\omegacdm$, $\log(10^{10} A_{\mathrm{s}})$, $\ln N_1$, $\ln N_2$, $\sigma_{\ln M| N^{\obs}_{\gal} }$, $\beta$, $\qnl$, $\sigma_{z}$, $B$, $\fnl$], with the cluster power spectrum.
 \item \textbf{CMB only}: seven free parameters [$\Omega_{\mathrm{b}}$, $h$, $\tau$, $n_{\mathrm{s}}$, $A_{\mathrm{sz}}$, $\omegacdm$, $\log(10^{10} A_{\mathrm{s}})$], with CMB data only.
 \item \textbf{CMB+clusters}: 14 free parameters [$\Omega_{\mathrm{b}}$, $h$, $\tau$, $n_{\mathrm{s}}$, $A_{\mathrm{sz}}$, $\omegacdm$, $\log(10^{10} A_{\mathrm{s}})$, $\ln N_1$, $\ln N_2$, $\sigma_{\ln M| N^{\obs}_{\gal} }$, $\beta$, $\qnl$, $\sigma_{z}$, $B$], with CMB and all cluster data.
 \item \textbf{CMB+clusters with \boldmath{$\fnl$}}: 15 free parameters [$\Omega_{\mathrm{b}}$, $h$, $\tau$, $n_{\mathrm{s}}$, $A_{\mathrm{sz}}$, $\omegacdm$, $\log(10^{10} A_{\mathrm{s}})$, $\ln N_1$, $\ln N_2$, $\sigma_{\ln M|N^{\obs}_{\gal} }$, $\beta$, $\qnl$, $\sigma_{z}$, $B$, $\fnl$], with CMB and all cluster data.
\end{enumerate}

\subsection{Results}
\label{sec:constraints}

\begin{figure}
\begin{center}
\includegraphics[width=0.9\linewidth]{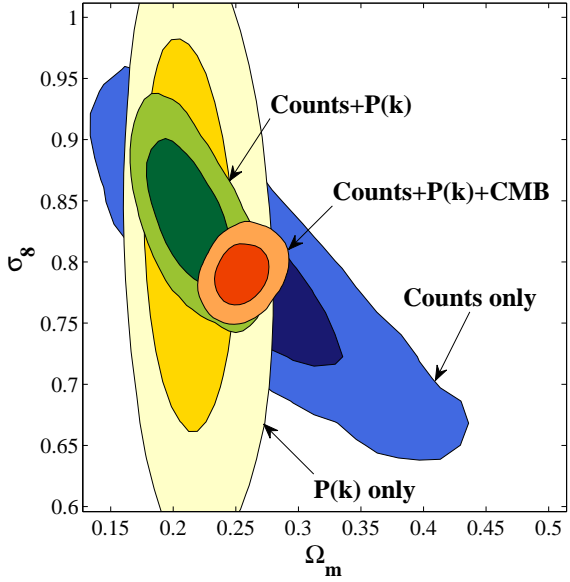}
\caption{Marginalized posterior probability distributions on the parameters $\omegam$--$\sigma_8$ for the runs using \textbf{Counts only} (blue), \textbf{Counts+\boldmath{$P(k)$}} (green) and \textbf{Counts+\boldmath{$P(k)$}+CMB} (orange), at 68 and 95 per cent confidence levels. The yellow contours show the joint constraints in the case of $P(k)$ data only.}
\label{fig:6par9parcmb_omsig8}
\end{center}
\end{figure}

\begin{figure}
\begin{center}
\includegraphics[width=0.9\linewidth]{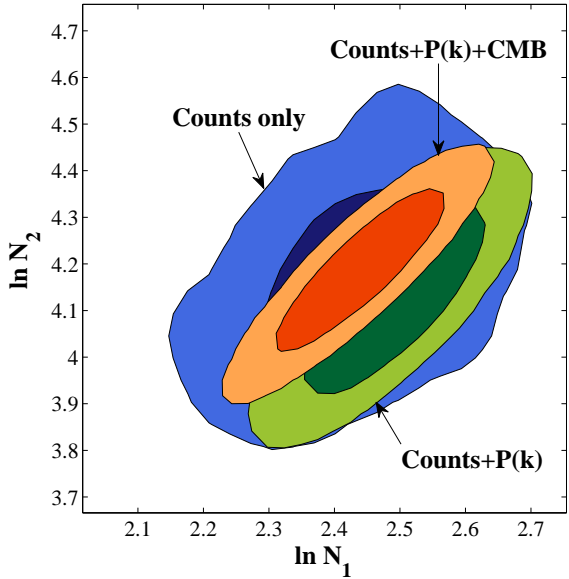}
\caption{Constraints on the scaling relation parameters for the runs using \textbf{Counts only} (blue), \textbf{Counts+\boldmath{$P(k)$}} (green) and \textbf{Counts+\boldmath{$P(k)$}+CMB} (orange), at 68 and 95 per cent confidence levels. Note that $\ln N_1 \equiv \ln N_{\gal}|M_1$ and $\ln N_2 \equiv \ln N_{\gal}|M_2$, where $M_1 = 1.3 \times 10^{14} \mbox{M}_{\bigodot}$ and $M_2 = 1.3 \times 10^{15}\mbox{M}_{\bigodot}$.}
\label{fig:6par9par_L1L2}
\end{center}
\end{figure}

\begin{figure}
\begin{center}
\includegraphics[width=0.9\linewidth]{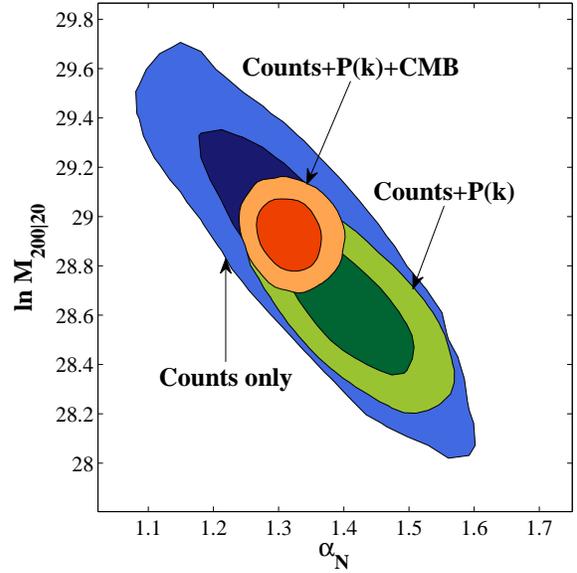}
\caption{Constraints on the slope $\alpha_N$ and intercept $\ln M_{200|20}$ of the scaling relation for the runs using \textbf{Counts only} (blue), \textbf{Counts+\boldmath{$P(k)$}} (green) and \textbf{Counts+\boldmath{$P(k)$}+CMB} (orange), at 68 and 95 per cent confidence levels.}
\label{fig:scalrel}
\end{center}
\end{figure}

\begin{figure}
\begin{center}
\includegraphics[width=0.9\linewidth]{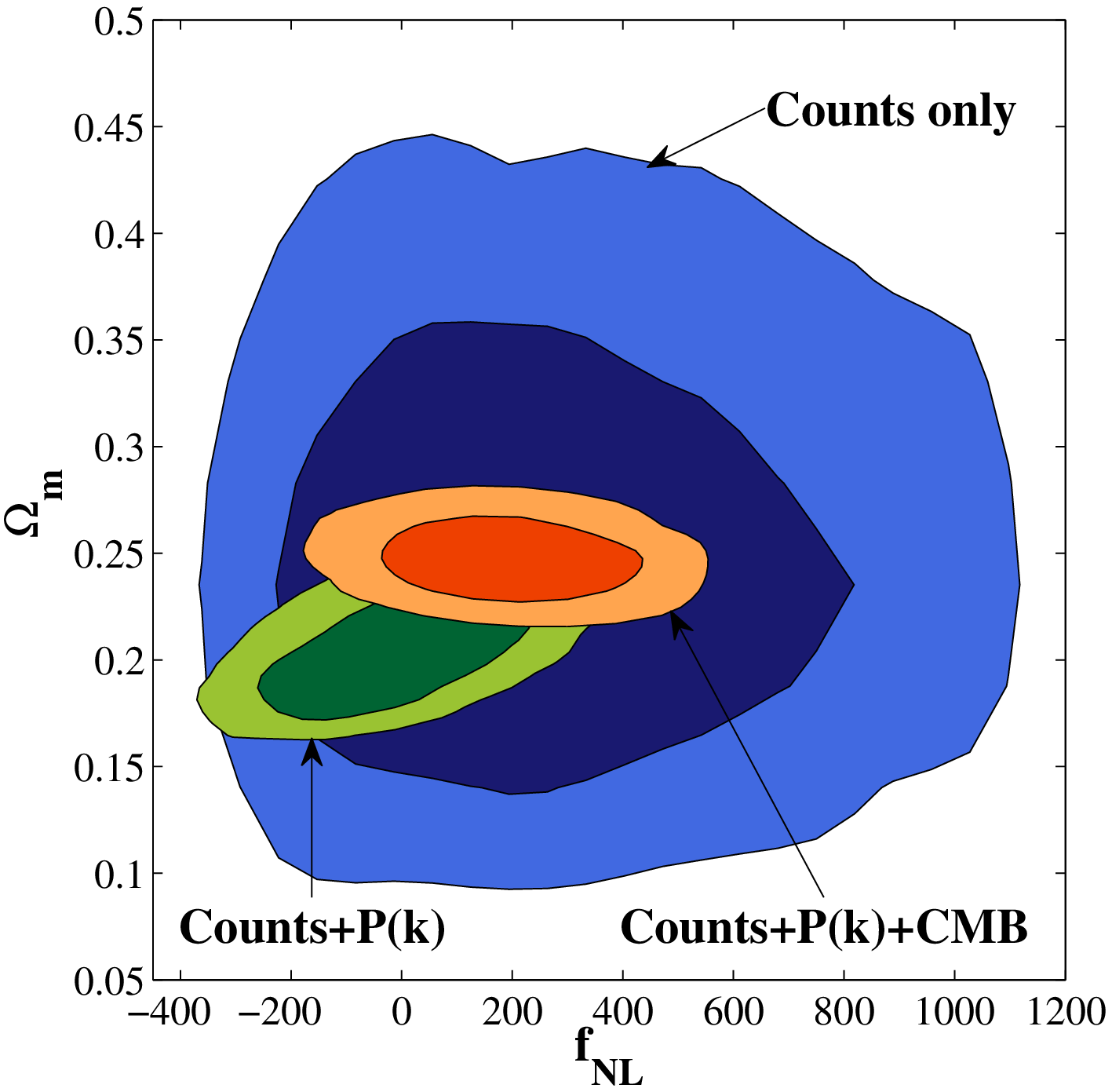}
\caption{Marginalized posterior probability distributions on the parameters $\fnl$--$\omegam$ for the runs using \textbf{Counts with \boldmath{$\fnl$}} (blue), \textbf{Counts+\boldmath{$P(k)$} with \boldmath{$\fnl$}} (green) and \textbf{CMB+clusters with \boldmath{$\fnl$}} (orange), at 68 and 95 per cent confidence levels. }
\label{fig:fnlom}
\end{center}
\end{figure}

\begin{figure}
\begin{center}
\includegraphics[width=0.9\linewidth]{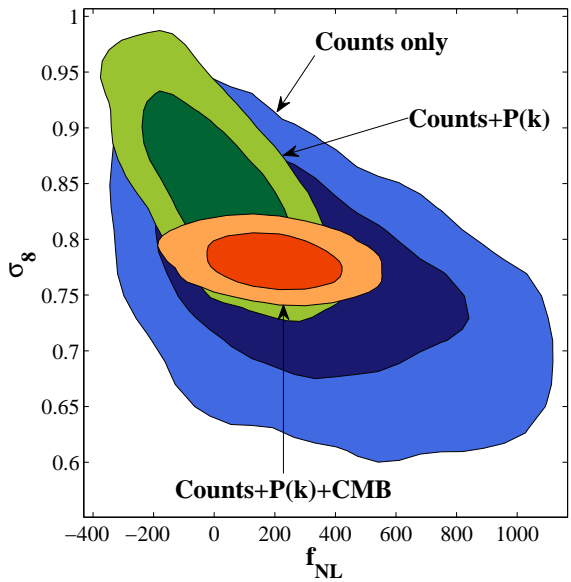}
\caption{Marginalized posterior probability distributions on the $\fnl$--$\sigma_{8}$ plane for the runs including \textbf{Counts with \boldmath{$\fnl$}} (blue), \textbf{Counts+\boldmath{$P(k)$} with \boldmath{$\fnl$}} (green) and \textbf{CMB+clusters with \boldmath{$\fnl$}} (orange), at 68 and 95 per cent confidence levels.}
\label{fig:fnlsig8}
\end{center}
\end{figure}

We summarize our results in Table \ref{tab:Table5results}, and we show in Figs \ref{fig:6par9parcmb_omsig8}--\ref{fig:fnlsig8} the 2D 68 and 95 per cent marginalized confidence regions for different pairs of parameters in our analysis. The colour scheme is the same for all figures: blue contours refer to runs with counts and WL mean masses data only, green contours include in addition the cluster power spectrum data, while orange contours also include CMB data from \emph{WMAP7}.

The joint constraint in the $\omegam$--$\sigma_{8}$ plane in Fig.~\ref{fig:6par9parcmb_omsig8} displays the typical degeneracy from cluster counts: the counts increase with increasing $\omegam$ and $\sigma_{8}$ values; hence, any increase in $\omegam$ must be balanced by a decrease in $\sigma_8$ (and vice versa), to keep the abundances at the observed values. The constraints on individual parameters with counts and masses only are consistent with \cite{Rozo2010}, and we find $\omegam = 0.25 \pm 0.06$ and $\sigma_8 = 0.80 \pm 0.06$ ($1\sigma$ errors throughout), while the errors are improved by a factor between 1.5 and 3, depending on the parameter, when adding the maxBCG power spectrum: in this case we obtain $\omegam = 0.215 \pm 0.022$ and $\sigma_8 = 0.84 \pm 0.04$. Combining then these results with the CMB data, the constraints shrink to $\omegam = 0.255 \pm 0.014$ and $\sigma_8 = 0.790 \pm 0.016$: the contribution of the CMB tightens the errors by a further factor of 2. As an interesting comparison, we show also the joint constraints for the case of $P(k)$ data only (yellow contours), with a prior on the scaling relation parameters: the degeneracy direction is complementary to that of the counts. The size of the posterior on $\sigma_8$ in this case depends on the assumptions made on the cluster bias: allowing for a completely free bias would cause a complete degeneracy with $\sigma_8$. Here the degeneracy is partially broken because we are instead assuming that the bias is centred around the predicted valued from the mass function, allowing only for deviations from this (parametrized by the scatter $B$), whose amplitude is limited by the prior on $B$.
 
In Fig.~\ref{fig:6par9par_L1L2}, we show the marginalized posterior probability contours of the scaling relation parameters $\ln N_1$ and $\ln N_2$. Constraints on individual parameters using counts and masses only are again compatible with \cite{Rozo2010} ($\ln N_1 = 2.44 \pm 0.11$, $\ln N_2 = 4.16 \pm 0.15$), while errors are reduced when adding the power spectrum, even if less significantly ($\ln N_1 = 2.49 \pm 0.09$, $\ln N_2 = 4.13 \pm 0.13$). Combining these results with the CMB data, the constraints are almost identical. 
Our constraints on the scaling relation scatter $\sigma_{\ln M}$ are in agreement with \cite{Rozo2010}, and they are not improved by the addition of power spectrum and CMB data.
We then calculated the likelihood contours on the derived parameters $\alpha_N$ and $\ln M_{200|20}$, which have a more direct physical interpretation as slope and intercept of the scaling relation (see equation~\ref{eq:scalingRelation}): these are shown in Fig.~\ref{fig:scalrel}. The marginalized mean values and $1\sigma$ errors on individual parameters, using counts and masses only, are $\alpha_N = 1.35 \pm 0.11$ and $\ln M_{200|20} = 28.85 \pm 0.33$. When adding the power spectrum, the errors reduce to $\alpha_N = 1.41 \pm 0.06$ and $\ln M_{200|20} = 28.64 \pm 0.17$. Combining then with the CMB data, the constraints are further improved to $\alpha_N = 1.32 \pm 0.03$ and $\ln M_{200|20} = 28.93 \pm 0.09$.

\begin{table*}
\begin{center}
\caption{Parameters used in the analysis and their assumed priors. In the prior columns a single number $n$ stands for a fixed value, $[a,b]$ stands for a flat prior, and $\mu \pm \sigma$ means a Gaussian prior of mean $\mu$ and standard deviation $\sigma$.}

    \begin{tabular} {c c c c c}
    \hline 
    Type & Symbol & Definition & Prior without CMB & Prior with CMB \\ [0.5ex] 
    \hline 
    Cosmology & $h$ & Dimensionless Hubble parameter & $0.7$ & $[0.4,0.9]$\\ 
    & $n_{\mathrm{s}}$ & Scalar spectral index  & $0.96$ & $[0.5,1.5]$\\ 
    & $\Omega_{\mathrm{b}}$ & Baryon energy density & $0.04397$ & $[0.01,0.2]$\\ 
    & $\Omega_{\mathrm{c}}$ & Cold dark matter energy density & $[0.1,0.9]$ & $[0.1,0.9]$\\
    & $\log(10^{10} A_{\mathrm{s}})$ & Amplitude of primordial perturbations & $[0.1,6.0]$ & $[0.1,6.0]$\\
    & $\tau$ & Optical depth & $0.09$ & $[0.01,0.125]$\\
    & $\fnl$ & Primordial non-Gaussianity amplitude & $[-900,900]$ & $[-900,900]$\\
    \hline 
    Scaling relation & $\ln N_1\equiv \ln N_{\gal}|M_1$ & Richness at $M_1=1.3\times10^{14}\mbox{M}_{\bigodot}$ & $[1.0,4.0]$ & $[1.0,4.0]$\\
    & $\ln N_2\equiv \ln N_{\gal}|M_2$ & Richness at $M_2=1.3\times10^{15}\mbox{M}_{\bigodot}$ & $[3.0,6.0]$ & $[3.0,6.0]$\\
    & $\sigma_{\ln M| N^{\obs}_{\gal} }$ & Scatter & $0.45\pm0.1$ & $0.45\pm0.1$\\
    \hline 
    Nuisance & $\beta$ & Weak-lensing mass measurements bias & $1.0\pm0.06$ & $1.0\pm0.06$\\ 
    & $B$ & Scatter on bias derived from mass function & $1.0\pm0.15$ & $1.0\pm0.15$\\
    & $\qnl$ & Non-linear correction to power spectrum & $[0.0,50.0]$ & $[0.0,50.0]$\\
    & $\sigma_{z}$ & Photo-$z$ errors & $[0,120]$ & $[0,120]$\\ 
    & $A_{\mathrm{SZ}}$ & Amplitude of CMB SZ template & $1$ & $[0,2]$\\
    \hline 
    Derived & $\omegam$ & Total matter energy density &  -- &  --\\ 
            & $\sigma_8$ & Amplitude of density perturbations &  -- &  --\\ 
    \hline 
    \end{tabular}   	
        
    \label{tab:Table4parameters} 
\end{center}
\end{table*}

\begin{table*}
\begin{center}
\caption{Marginalized mean values and $1\sigma$ errors on the cosmological parameters, for the runs \textbf{Counts only}, \textbf{Counts with \boldmath{$\fnl$}}, \textbf{Counts+\boldmath{$P(k)$}}, \textbf{Counts+\boldmath{$P(k)$} with \boldmath{$\fnl$}}, \textbf{CMB+clusters} and \textbf{CMB+clusters with \boldmath{$\fnl$}}. Note that $\omegam$ and $\sigma_8$ are derived parameters in our analysis.}
	\centering
	\begin{tabular} {c c c c c c c}
		\hline
		\vspace{-0.3cm}\\ 
    Parameters & \multicolumn{2}{c}{{\textbf{Counts only}}} & \multicolumn{2}{c}{\textbf{Counts+\boldmath{$P(k)$}}} & \multicolumn{2}{c}{\textbf{Clusters+CMB}} \\ [0.5ex]
    \cline{2-3} \cline{4-5} \cline{6-7}
           & no $\fnl$ & +$\fnl$ & no $\fnl$ & +$\fnl$ & no $\fnl$ & +$\fnl$ \\
    \hline\\
    $\omegam$ & $0.25\pm0.06$ & $0.25\pm0.06$ & $0.215\pm0.022$ & $0.209\pm0.022$ & $0.255\pm0.014$ & $0.248\pm0.013$ \\
    $\sigma_{8}$ & $0.80\pm0.06$ & $0.77\pm0.07$ & $0.84\pm0.04$ & $0.85\pm0.05$ & $0.790\pm0.016$ & $0.780\pm0.016$ \\
    $\ln N_{1}$ & $2.44\pm0.11$ & $2.44\pm0.11$& $2.49\pm0.09$ & $2.49\pm0.08$ & $2.44\pm0.08$ & $2.43\pm0.08$ \\
    $\ln N_{2}$ & $4.16\pm0.15$ & $4.15\pm0.15$& $4.13\pm0.13$ & $4.11\pm0.12$ & $4.19\pm0.11$ & $4.15\pm0.11$ \\
    $\sigma_{\ln M}$ & $0.38\pm0.06$ & $0.38\pm0.06$ & $0.36\pm0.06$ & $0.37\pm0.06$ & $0.378\pm0.059$ & $0.38\pm0.06$ \\
    $\beta$ & $1.00\pm0.06$ & $1.01\pm0.06$ & $1.01\pm0.06$ & $1.01\pm0.06$ & $1.01\pm0.06$ & $1.00\pm0.06$ \\
    $\qnl$ & -- & -- & $26\pm10$ & $27\pm10$ & $14\pm6$ & $16 \pm 7$ \\ 
    $\sigma_{z}$ & -- & -- & $46\pm12$ & $42 \pm 8$ & $43\pm10$ & $31\pm5$ \\
    $B$ & -- & -- & $1.07\pm0.13$ & $1.01\pm0.15$ & $1.19\pm0.11$ & $1.00\pm0.14$ \\
    $\fnl$ & -- & $282 \pm 317$ & -- & $12\pm157$ & -- & $194\pm128$ \\\\[0.5ex]
    \hline 
   \end{tabular}   	
        \label{tab:Table5results}
\end{center}
\end{table*}


Figs \ref{fig:fnlom} and \ref{fig:fnlsig8} show the constraints on $\fnl$ and its degeneracies with $\omegam$ and $\sigma_8$. First we can see that, when only counts and masses are used, the constraints on $\fnl$ are weak as expected. The situation improves when adding the cluster power spectrum: in this case, the constraints are tighter, and we observe a positive correlation between $\fnl$ and $\omegam$ and an anti-correlation with $\sigma_8$.
In fact, if we increase (decrease) $\omegam$, the peak of the power spectrum decreases (increases) while also being shifted to higher (lower) values of $k$, while $\sigma_{8}$ simply changes the overall normalization.
As described above, an increase in $\fnl$ causes a boost in the power spectrum on large scales (small $k$), so that $\sigma_8$ needs to decrease to compensate a higher $\fnl$: this is exactly what is shown in Fig.~\ref{fig:fnlsig8}.
In addition to this, $\omegam$ should increase to compensate a higher $\fnl$: this can be seen in Fig.~\ref{fig:fnlom}.
We also see that the addition of the CMB power spectrum data improves the constraints on $\omegam$ and $\sigma_8$ and only indirectly reduces the bounds on $\fnl$, since PNG simply affects the higher-order statistics of the CMB.\\

Our constraints on PNG are $\fnl = 12 \pm 157 $ ($1\sigma$) (without CMB) and $\fnl = 194 \pm 128 $ (with CMB), which are statistically compatible with zero and with each other. The shift in the mean between the two results is clear by looking at Figs \ref{fig:fnlom} and \ref{fig:fnlsig8}: the addition of the CMB favours lower values of $\sigma_8$ (and higher values of $\omegam$), thus shifting the favoured $\fnl$ values in the process.
While not competitive with results from the CMB bispectrum or from combined analyses of multiple galaxy surveys, it is interesting to find such constraints independently and for the first time with the clustering of galaxy clusters.\\

Since we restrict our analysis to nearly linear scales, by imposing the data cut at $k_{\max}=0.15 \, h$ Mpc$^{-1}$, we are  not expecting strong constraints on $\qnl$. The constraints we found are indeed broad and in agreement within the errors with the results by \citet{Huetsi2010}, who found $\qnl = 14.2 \pm 2.8$ when marginalizing over three parameters only: we obtain $\qnl = 26 \pm 10$ and $\qnl = 14 \pm 6$ when also using CMB data.\\

It is also worth mentioning the results on the $B$ parameter, which was introduced to take into account the uncertainty in the bias expression derived from the mass function. As this parameter allows an arbitrary constant rescaling of the bias, it also has the desirable property of cancelling the effect of the scale-independent bias correction $\delta b (\fnl)$, as described in Section~\ref{sec:fnl}.
To check that the Gaussian prior we are imposing $B=1.0\pm0.15$ is large enough for both purposes, we made an additional run replacing it with a flat prior $B \in [0.0001,5]$. In this way, we obtain nearly unchanged results on $\fnl$.

\section{Conclusions}
\label{sec:concl}

In this work, we have investigated the cosmological implications of the optically selected SDSS maxBCG galaxy cluster data, obtaining extended cosmological constraints with respect to previous works. We considered the number counts of clusters in richness bins and the WL mass estimations, including the respective covariances, for a cross-calibration of the scaling relation. We then combined such data for the first time  with a measurement of the redshift-space power spectrum of the same clusters. In the modelling we included an effective treatment of the non-linear contribution, photo-$z$ smoothing, RSDs and Alcock--Paczynski effect. We only considered quasi-linear scales at $k < k_{\max} = 0.15 \, h$ Mpc$^{-1}$ to be conservative.
 We estimated the off-diagonal terms of the counts-clustering covariance matrix with a jackknife method applied on both  data and $N$-body simulations, and found consistently that such contributions are negligible.

We then performed a full MCMC analysis of the posterior probability distribution of cosmological parameters given the full data set.
 By thus combining the one- and two-point statistics, we achieved a factor 1.5--3 improvement on the errors on the cosmological parameters, if compared with previous analyses using number counts and masses only \citep{Rozo2010}, obtaining e.g. for the fluctuation amplitude $  \sigma_8 = 0.84 \pm 0.04 $ ($1\sigma$) and for the matter content $\omegam = 0.215 \pm 0.022$ ($1\sigma$). These are further tightened by a factor of 2 by the addition of the CMB data. On the other hand, we found that the errors on the scaling relation parameters are consistent with previous works, but not significantly improved by the addition of the cluster power spectrum.

 As an interesting application, we also tested PNG, which is constrained through the non-Gaussian halo mass function and the scale-dependent cluster bias. Assuming deviations from Gaussianity at the three-point (skewness) level of the local type, we obtained $\fnl = 12 \pm 157$ ($1\sigma$) from our combined data set, which shifts to $\fnl = 194 \pm 128$ ($1\sigma$) when including the \emph{WMAP7} CMB data. While not competitive with the CMB bispectrum and with results from combined galaxy clustering data sets, this result is consistent with them and was obtained using the maxBCG cluster data alone. Our results can be seen as a proof of concept towards a full joint analysis of the LSS,  consistently including both galaxies and clusters as dark matter tracers, to achieve the full potential of the upcoming galaxy surveys such as the Dark Energy Survey and the \emph{Euclid} mission.
%
%

\section*{Acknowledgements}
We are grateful to Ixandra Achitouv, Richard Battye and Eduardo Rozo for useful discussions.
We acknowledge support from the Trans-Regional Collaborative Research Centre TRR 33 -- 'The Dark Universe of the Deutsche Forschungsgemeinschaft (DFG)'. 

\bibliographystyle{mnras}
\bibliography{ms}

\label{lastpage}
\end{document}